\documentclass[prl,showpacs,preprintnumbers,twocolumn]{revtex4}
\usepackage{amssymb}
\usepackage{color}
\usepackage{graphicx}

\begin{document}

\title[]{Asymmetric metal-insulator transition in disordered ferromagnetic films}
\author{R. Misra, A.F. Hebard}
\email[Corresponding author:~]{afh@phys.ufl.edu}
\author{K.A. Muttalib}
\affiliation{Department of Physics, University of Florida, Gainesville FL 32611-8440}
\author{P. W\"{o}lfle}
\affiliation{Institute for Condensed Matter Theory, Institute for Nanotechnology and 
Center for Functional Nanostructures D-76128 Karlsruhe, Germany}
\author{}

\begin{abstract}
We present experimental data and a theoretical interpretation on the
conductance near the metal-insulator transition in thin ferromagnetic Gd films of
thickness $b\approx 2-10nm$. A large phase relaxation rate caused by
scattering of quasiparticles off spin wave excitations renders the dephasing
length $L_{\phi }\lesssim b$ in the range of sheet resistances considered,
so that the effective dimension is $d=3$.  The observed approximate fractional
temperature power law of the conductivity is ascribed to the scaling regime
near the transition. The conductivity data as a function of temperature and
disorder strength collapse on to two scaling curves for the metallic and
insulating regimes. The best fit is obtained for a dynamical exponent 
$z\approx 2.5$ and a correlation length critical exponent $\nu'\approx 1.4$ on the 
metallic side and a localization length exponent $\nu\approx 0.8$ on the insulating side. 
\end{abstract}

\maketitle
\date{June 4, 2009}

Since the scaling theory of Anderson localization has been proposed in 1979 
\cite{aalr}, the metal-insulator transition in disordered conductors \cite%
{anderson} has been one of the most extensively studied cases of quantum
phase transition, both experimentally and theoretically. In its simplest
form it describes non-interacting electrons in a disordered potential, where
the disorder can be controlled experimentally in a variety of ways, e.g by
systematic doping. One of the most dramatic predictions of the scaling
theory is the absence of extended states, and therefore true metallic
behavior, in systems in dimensions $d\le 2$. This has been verified in numerous experiments 
\cite{review1}. The other prediction is the existence of a critical point in $%
d>2$ where the conductivity in the metallic phase goes to zero continuously
with increasing disorder, in contrast to having a minimum metallic
conductivity \cite{mott}. Electron-electron interactions are known to modify
the behavior near a metal-insulator transition in a significant way \cite{review2}.
For example, indications of a metallic state in two-dimensional systems have
been found in experiment, and a number of theoretical scenarios explaining
such a state have been developed \cite{review3}.

Near the transition, the behavior is characterized by power laws with
critical exponents. For example, the dc conductivity $\sigma (\lambda )$,
with $\lambda $ being a measure of disorder, follows a power law $\sigma
\sim t^{s}$ , where $t=(1-\lambda /\lambda _{c})$ denotes the distance to
the critical point at the critical disorder $\lambda _{c}$ and $s$ is the
conductivity exponent. The dynamical conductivity at the critical point, on
the other hand, is characterized by the dynamical exponent $z$  as 
$\sigma (\omega ;\lambda _{c})\sim \omega ^{1/z}$. The correlation
length on the metallic side ($\lambda < \lambda_c$)
diverges at the critical point as $\ \xi \sim t^{-\nu' }$ and the localization 
length ($\lambda > \lambda_c$) diverges as $\xi \sim |t|^{-\nu}$. The critical exponents
$\nu$ and $\nu'$ may be different. 
In $d=3$ dimensions the relation $s=\nu' $ holds. The
exponents in $d=3$ have not been calculated in a reliable way up to now. 

As for any quantum phase transition, the critical exponents can not be
measured experimentally at the true $T=0$ critical point, but must be
inferred from finite temperature $T$ measurements. Therefore, the emphasis
has been to obtain the conductivity as a function of $T$, as close to $T=0$
as possible. In spite of intense efforts over several decades \cite{stupp}
it turned out to be rather difficult to access the critical regime in a
reliable way. While so far all such experiments confirm the continuous
nature of the transition, the values of the critical exponents remain
controversial. Published experimental values of $s$ and $z$ vary from $%
s\approx 0.5$ \cite{paanalen}, $s\approx 1$ \cite{field} to $s\approx 1.6$
and from $z\approx 2$ \cite{bogdanovich} to $z\approx 2.94$ \cite{waffen}.
For the Anderson transition (omitting interaction effects), numerical studies 
find the conductivity exponent $s\approx 1.6$ \cite{slevin} while theoretical 
prediction for the dynamical exponent is $z=3$  \cite{wegner}.

Theoretically, the scale dependent conductivity at finite $T$ is obtained
from $\sigma (\omega )$ by replacing $\omega $ by $T$. The critical
dynamical scaling is found even away from the
critical point, at frequencies $\omega >\omega _{\xi }$, where $\omega _{\xi
}=\frac{1}{\tau }(\xi /l)^{-z}$. Here $l$ and $\tau $ are the mean free path
and the momentum relaxation time, respectively. More precisely, in the above
scaling regime the dynamical conductivity obeys the scaling law,
\begin{equation}\label{G}
\sigma (\omega ;\lambda )=\xi ^{-1}G(\pm 1,\xi \omega
^{1/z}),\;\;\;t\gtrless 0.
\end{equation}%
At the critical point, when $\xi \rightarrow \infty $, it follows that 
$G(\pm 1,\xi \omega^{1/z})\sim \xi \omega^{1/z}$. Using the
sheet resistance $R_{0}$ as the disorder parameter controlled in experiment, 
so that $\xi\propto |R_0-R_c|^{-\nu}$, and replacing $\omega$ by $T$, 
the conductivity should obey the scaling,
\begin{equation}\label{scaling}
|R_{0}-R_{c}|^{-\nu }\sigma (T;R_{0})=G(\pm 1,|R_{0}-R_{c}|^{-\nu
}T^{1/z})
\end{equation}%
with $(R_{0}-R_{c})\gtrless 0$, where $R_{c}$ is the critical resistance. Below we 
will see that the exponent $\nu$ can be different on the two sides of the transition.

In this letter we report the study of the conductivity near the metal-insulator
transition in a thin-film geometry where it is possible not only to
increase the disorder directly by changing the atomic structure, e.g. by varying
the film deposition parameters and by annealing, but also indirectly by
varying the film thickness. Both factors determine the sheet resistance,
which is the single important parameter controlling the distance to the
critical point. We are able to measure the disorder as well as the temperature
dependence of the conductivity quite reliably and reproducibly. In
particular, we are able to determine the location of the critical point with
unprecedented precision. As we will explain below, at the relatively high
temperatures and large sheet resistances of the experiment, the films are in
an effectively three-dimensional regime.

Two series of thin films of Gd (series~1 and series~2) were grown by r.f.
magnetron sputtering through a shadow mask onto sapphire substrates held at
a temperature of 130\thinspace K. The current and voltage leads of the
deposited sample overlapped with predeposited palladium contacts, thus
allowing reliable electrical connection with low contact resistance for 
\textit{in situ} measurements of the electrical properties. The experiments
were performed in a specialized apparatus in which the sample can be
transferred without exposure to air from the high vacuum deposition chamber
to an adjoining low-temperature cryostat and electrically reconnected for
transport measurements. Immediately after deposition, the samples were
transferred to the cryostat and held at a temperature of 77\thinspace K or
below. At these temperatures the samples are stable and do not undergo any
time-dependent changes in resistance. If however the temperature is
temporarily raised back to the deposition temperature (130\thinspace K),
annealing marked by a slow irreversible increase in resistance occurs. 

To parameterize the amount of disorder in a given film\cite{mitra}, we use the
sheet resistance $R_{0}\equiv R_{xx}(T=5~K)$ where $R_{xx}$ is the
longitudinal resistance. In our experiments $R_{0}$ spans the range from $%
4~k\Omega $ (35 {\AA } thick) to 40~k$\Omega $ ($<$ 20 {%
\AA } thick). Controlled thermal annealing thus allows us to advantageously
tune a single sample through successive stages of increased disorder. Our
series~1 samples comprise 5 separate depositions with two of the samples
undergoing 12 successive anneals thus giving a total of 17 measurements at
different stages of disorder. Our series~2 samples comprise a single sample
undergoing 15 successive anneals for a total of 16 measurements spanning the
critical region where the metal-insulator transition occurs. The resistance
was measured using four-terminal dc techniques.

We now argue why the films are effectively three dimensional. As discussed in
Ref.~[\onlinecite{mitra}], the phase relaxation rate in ferromagnetic films
is dominated by the scattering off spin wave excitations, yielding
$\tau _{\varphi }^{-1}\propto T$, in $d=2,3$ dimensions.  The effective
dimensionality of the system is $d=3$ if the temperature dependent
correlation length $L_{\varphi }(T)=\sqrt{\sigma \tau _{\varphi }/N_{0}}\ll b
$, where $N_{0}$ is the $3d$ density of states. The inequality is satisfied
for temperatures $T\gg T_{x}=T_{0}(bk_{F})^{-3}\beta ^{3/2}(\epsilon
_{F}\tau _{\varphi ,0}/\hbar )$ , where $\epsilon _{F},k_{F}$ are Fermi
energy and wave number, respectively, \ and $1/\tau _{\varphi ,0}\sim 1K$ is
the phase relaxation rate at $T=T_{0}=1K$ and $\beta $ is a coefficient of
order unity. For not too thin samples, $bk_{F}>10$ , the $3d$ condition is
seen to be met in our experiments.

\begin{figure}[tbp] 
\begin{center}
\includegraphics[angle=0, width=0.38\textheight]{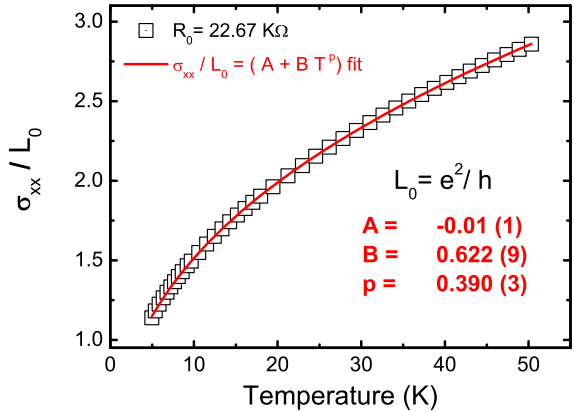}
\end{center}
\caption{Conductivity as a function of temperature for the sample with critical resistance $R_c=22.67 k\Omega$ closest to critical disorder where $A = 0$. The solid red line is a fit using Eq.~(\ref{fit}). The parameter $A$ has positive (negative) values for samples with $R_0 < R_c$ ($R_0 > R_c$)}
\label{Rcritical}
\end{figure}

To gain a first impression of the critical behavior exhibited by the data we
use a fit to the expression 
\begin{equation}\label{fit}
\sigma (T;R_{0})/L_0=BT^{p}+A.
\end{equation}
Here $L_0=e^2/h$ is the conductance quantum. The functional form of the conductivity given by Eq.~(\ref{fit}) provides a good
description of all our data for fourteen samples in the range $15 ~k\Omega < R_0< 31 ~k\Omega$.
We fit the data to Eq.~(\ref{fit}) allowing the parameters $p$, $B$ and $A$
to vary with disorder. Figure~\ref{Rcritical} shows the fit for the sample with a value of $R_{0}= 22.67~k\Omega$ closest to critical disorder $R_c$ where $A = 0$. We note that a positive (negative) value of $A$ indicates the delocalized (localized) regime. It follows from the scaling relation Eq.~(\ref{G}) that $A=0$ at the critical point.  Fortunately, the sample depicted in Fig.~\ref{Rcritical} with $R_{0}=22.67 ~k\Omega$ and $A=-0.01(1)$ is close enough to criticality so that we can confidently make the identification $R_{c}=22.67 ~k\Omega$. We note that $R_{c}$ is of order unity in units of 
$h/e^{2}$. Using the same fitting procedures for the thirteen samples away from criticality (i.e., $A \neq 0$) we allowed all three parameters, $A, B, p$  to vary with $R_{0}$. It turns out that $B$ and $p$ do vary with disorder, if only slightly. The scaling property Eq.(\ref{G}) requires the parameters $B, p$ to be independent of disorder. At criticality shown in Fig.~\ref{Rcritical}, we determine the parameter values (see inset) $p=0.390(3)$ and $B=0.622(9) K^{-p}$. We infer from the value of $p$ a critical dynamical exponent $z = 1/p \approx 2.5$. 

\begin{figure}
\begin{center}
\includegraphics[angle=0, width=0.35\textheight]{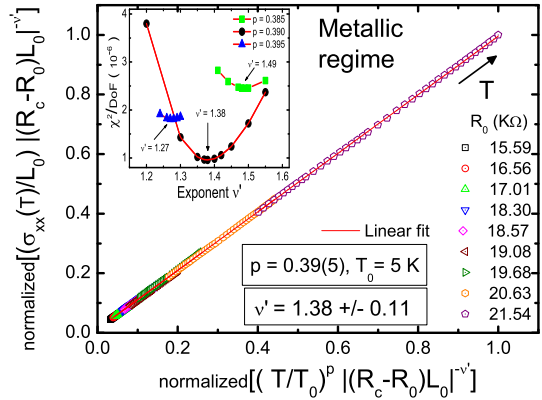}
\includegraphics[angle=0, width=0.35\textheight]{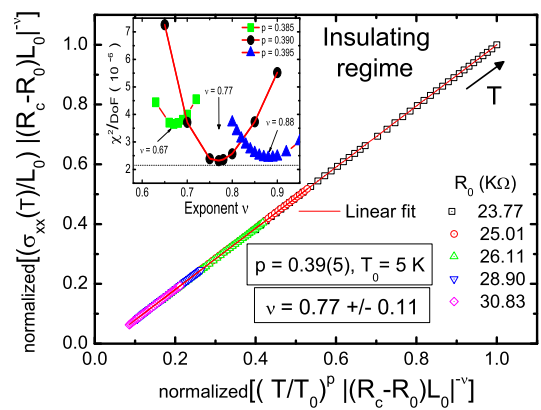}
\end{center}
\caption{Collapse of the data for different values of $R_0$ according to the 
scaling relation given by Eq.~(\ref{scaling}) for $\nu'=1.38$ on the metallic 
side (top panel) and $\nu=0.77$ 
on the insulating side (bottom panel). Here $T_0 = 5~K$ is a reference temperature. The insets of each panel show the dependence of $\chi^2$ on $\nu'$ and $\nu$ for the values of $p$ listed in the legends.  The best-fit values occur at $p=0.39$ with well-defined minima at $\nu'=1.38(11)$ and $\nu=0.77(11)$.}.
\label{collapse}
\end{figure}

Next we analyse the data in a completely unbiased way by using a
scaling plot of the scaling function $G$ defined in Eq.~(\ref{scaling}). For practical
reasons this requires picking a value of the critical disorder,  $R_{c}$,
which we take from the above analysis as $R_{c}=22.67 ~k\Omega $. We may now
take different values of the critical exponents $\nu' $ and $\nu$ to see how well the
data points collapse onto single curves on both sides of the transition. The
best data collapse is found for $\nu' = 1.38$ on the metallic side and 
$\nu = 0.77$ on the insulating side. The two panels of Fig.~\ref{collapse}
show how well the temperature-dependent data for each of the samples indicated in the legends collapse onto linear scaling curves for the two best fit values of $\nu'$ and $\nu$. For convenience, we have normalized the axes of each panel to unity using the highest temperature value of the conductivity of the samples $R_{0}=21.54 ~k\Omega$ ($R_{0}=23.77 ~k\Omega$) closest to criticality on the metallic (insulating) side of the transition.   

In the insets to the two panels of Fig.~\ref{collapse}, we show the dependence of Chi-square ($\chi^2$) on $\nu'$ in the metallic regime and $\nu$ in the insulating regime. The $\chi^2$ calculation is performed using statistical weights proportional to the inverse of the ordinate values, thereby increasing the sensitivity of $\chi^2$ to the data sets close to the origin and hence further away from criticality. The minima of $\chi^2$ are clearly defined for $p = 0.39$ at 
$\nu'=1.38(11)$ in the metallic regime and $\nu=0.77(11)$ in the insulating regime. Accordingly, the conductivity exponent $s$ in 3d has been experimentally determined to have the value $s=\nu'= 1.38(11)$. 

It is remarkable and somewhat surprising that the scaling functions are very well represented by straight lines in the regime considered. Thus the conductivity on the metallic side can be represented by an expression reminiscent of Eq.~(\ref{fit}) with, however, fixed coefficients $B, p=1/z$, $\sigma (T;R_{0})/L_0=BT^{1/z}+a|R_0-R_c|^{\nu'}$, and a disorder independent coefficient $a$ determined from the intercepts of the linear fits in Fig.~\ref{collapse}.

\begin{figure}[tbp]
\begin{center}
\includegraphics[angle=0, width=0.38\textheight]{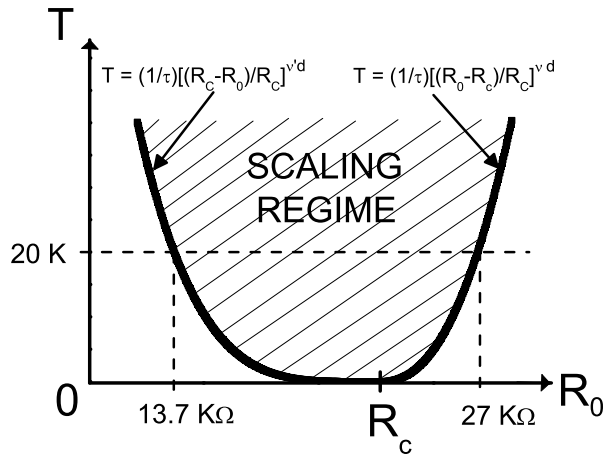}
\end{center}
\caption{Estimate of the width of the critical regime as a function of temperature, using the
experimentally obtained values of the exponents $\nu'$ and $\nu$.}
\label{width}
\end{figure}

As pointed out in Ref.~[\onlinecite{stupp}], reasons for the earlier
experiments not agreeing with one another have been traced to difficulties
in having a system allowing sufficient access into the critical region and
possessing a well defined critical point. In contrast, the current work is
done on a system where the critical point can be clearly identified, and the
critical region is experimentally accessible. Note that the number of data sets (14 total) around $R_c$ to be kept in the scaling analysis of Fig.~\ref{collapse} were determined by comparing the $\chi^2$ fits for different number of data points kept. A minimum in $\chi^2$ was obtained when the data points were restricted from about $15 k\Omega - 31 k\Omega$. We can also theoretically estimate the width of the critical 
region from the boundary frequency,
$\omega_{\xi}=\frac{1}{\tau}(\xi/l)^{-z}\sim \frac{1}{\tau} (|R_0-R_c|/R_c)^{\tilde{z}}$,
where $\tilde{z}$ is equal to $\nu' z$ on the metallic and $\nu z$ on the insulating side. At $T\sim 20 K$ and using $1/\tau\sim 10^{3}K$, 
this gives $|R_0-R_c|/R_c\sim 0.4$ on the metallic side and $\sim 0.2$ on the insulating side, corresponding to a critical region extending from about $15 ~k\Omega - 28 ~k\Omega$, as shown in Fig.~\ref{width}. This is a sufficiently large experimentally accessible region that allows us to obtain the critical exponents quite reliably.

Although the annealing treatment of a sample in the critical regime allows
one to move the system through the transition point in small steps, it is
desirable to have in addition a continuous control parameter. We have observed that
an applied magnetic field $H$ shifts the
parameter $A$ smoothly. A continuous variation of $H$ therefore should allow to
access the critical behavior of $A$ in the range one order of magnitude smaller
than the present one.
Work in this direction is in progress.

Our finding of the two distinctly different values for the critical exponent $\nu$ in the metallic and insulating phase, respectively, is unexpected. The quality of the data and of the fit to the scaling function appears to be so good that the difference of $\nu$ and $\nu'$ can not be explained by uncertainties in the measurement or deviations from the scaling form. From the point of view of theory, different critical exponents of the correlation length on both sides of the transition may indicate a different structure of critical modes. We can only speculate that this behavior may be related to the system considered here to be ferromagnetic. In ferromagnets we expect spin transport by spin wave propagation, which should render it different from charge transport. To our knowledge a theoretical treatment of this problem is not available at present.

To summarize, we have studied the metal-insulator transition in thin
ferromagnetic disordered films in an effectively three-dimensional regime
experimentally and theoretically. We concentrate on the critical regime, for
which the dynamical scaling behavior at the critical point is known to be $%
\sigma (\omega )\propto \omega ^{1/z}$. At finite temperatures the relevant
frequencies $\omega $ are given by the temperature $T$.  A first fit of the 
$\sigma (T)$ data to a temperature power law plus a $T$-independent constant 
$A(R_{0})$ showed an exponent $p\approx 0.4$ and serves to determine the
critical resistance $R_{c}$ from the zero of $A(R_{0})$. A full scaling
analysis of the data from a reasonably accessible  regime around the critical point ($\pm
30\%$) allows to determine the critical exponent of the correlation length
as $\nu' = 1.38(11) $ on the metallic side and the localization length exponent 
$\nu = 0.77(11)$ on the insulating side and the dynamical critical exponent as $z\approx
2.5$.

We acknowledge discussions with E. Abrahams, F. Evers, A. Finkel'stein, Y. Imry. and D.
Khmelnitskii. This work has been supported by the NSF under
Grant No. 0704240 (AFH), and by the DFG-Center for Functional Nanostructures
(KAM, PW).

\end{document}